\documentclass[referee]{raa}
\usepackage{graphicx,times}             
\usepackage{natbib}
\usepackage{amssymb,amsmath}
\usepackage{multirow}
\usepackage{booktabs}
\usepackage{color}
\usepackage{ulem}
\usepackage{threeparttable}
\bibpunct{(}{)}{;}{a}{}{,}
\usepackage{xcolor}
\newcommand{\TCDR}{Technical Report}
\newcommand{\ICDR}{Technical Report}
\newcommand{\rmxaa}{Revista Mexicana de Astronomía y Astrofísica}
\usepackage[colorlinks=true,linkcolor=green,anchorcolor=red,citecolor=blue,filecolor=red,urlcolor=red]{hyperref}
\begin{document}
   \title{Site-testing at the Muztagh-ata Site.V. Nighttime Cloud Amount during the Last Five Years
}

   \volnopage{Vol.0 (20xx) No.0, 000--000}      
   \setcounter{page}{1}          

   \author{Jing Xu
      \inst{1}
   \and Guo-jie Feng
      \inst{1}
	\and Guang-xin Pu
		\inst{1}
	\and Le-tian Wang
		\inst{1}
   \and Zi-Huang Cao
		\inst{2}
	\and Li-Qing Ren
		\inst{3}
	\and Xuan Zhang
		\inst{1}
	\and Shu-guo Ma
		\inst{1}
	\and Chun-hai Bai
		\inst{1}
   \and Ali Esamdin
      \inst{1}
   \and Jian Li
		\inst{2}
   \and Yuan Tian
		\inst{2}
   \and Zheng Wang
		\inst{2}
   \and Yong-heng Zhao
		\inst{2}
   \and Jian-rong Shi
		\inst{2}
}

   \institute{Xinjiang Astronomical Observatory, Chinese Academy of Sciences, 
				Urumqi, 830011, China; {\it xujing@xao.ac.cn,fengguojie@xao.ac.cn}\\
        \and
             National Astronomical Observatories, Chinese Academy of Sciences,
             Beijing 100012, China\\
 		\and
             Urumqi Meteorological Satellite Ground Station, 
             Urumqi, Xinjiang 830011, China\\
\vs\no
   {\small Received~~20xx month day; accepted~~20xx~~month day}}

\abstract{The clarity of nights is the major factor that should be carefully considered for optical/infrared astronomical observatories in site-testing campaigns. Cloud coverage is directly related to the amount of time available for scientific observations at observatories. In this article, we report on the results of detailed night-time cloud statistics and continuous observing time derived from ground-based all-sky cameras at the Muztagh-ata site from 2017 to 2021. Results obtained from acquisition data show that the proportion of the annual observing time at the Muztagh-ata site is 65\%, and the best period with the least cloud coverage and longer continuous observing time is from September to February. We made a comparison of the monthly mean observing nights obtained from our all-sky cameras and CLARA dataset, results show that the discrepancy between them may depend on the cloud top heights. On average, this site can provide 175 clear nights and 169 nights with at least 4 hours of continuous observing time per year.
\keywords{atmospheric effects;site-testing; cloud amount; all-sky camera}
}

   \authorrunning{J. Xu et al}            
   \titlerunning{Site-testing at the Muztagh-ata Site.V.}  

   \maketitle


%

\section{Introduction}           
\label{sect:intro}
\begin{figure}
   \centering
  \includegraphics[width=10cm, angle=0]{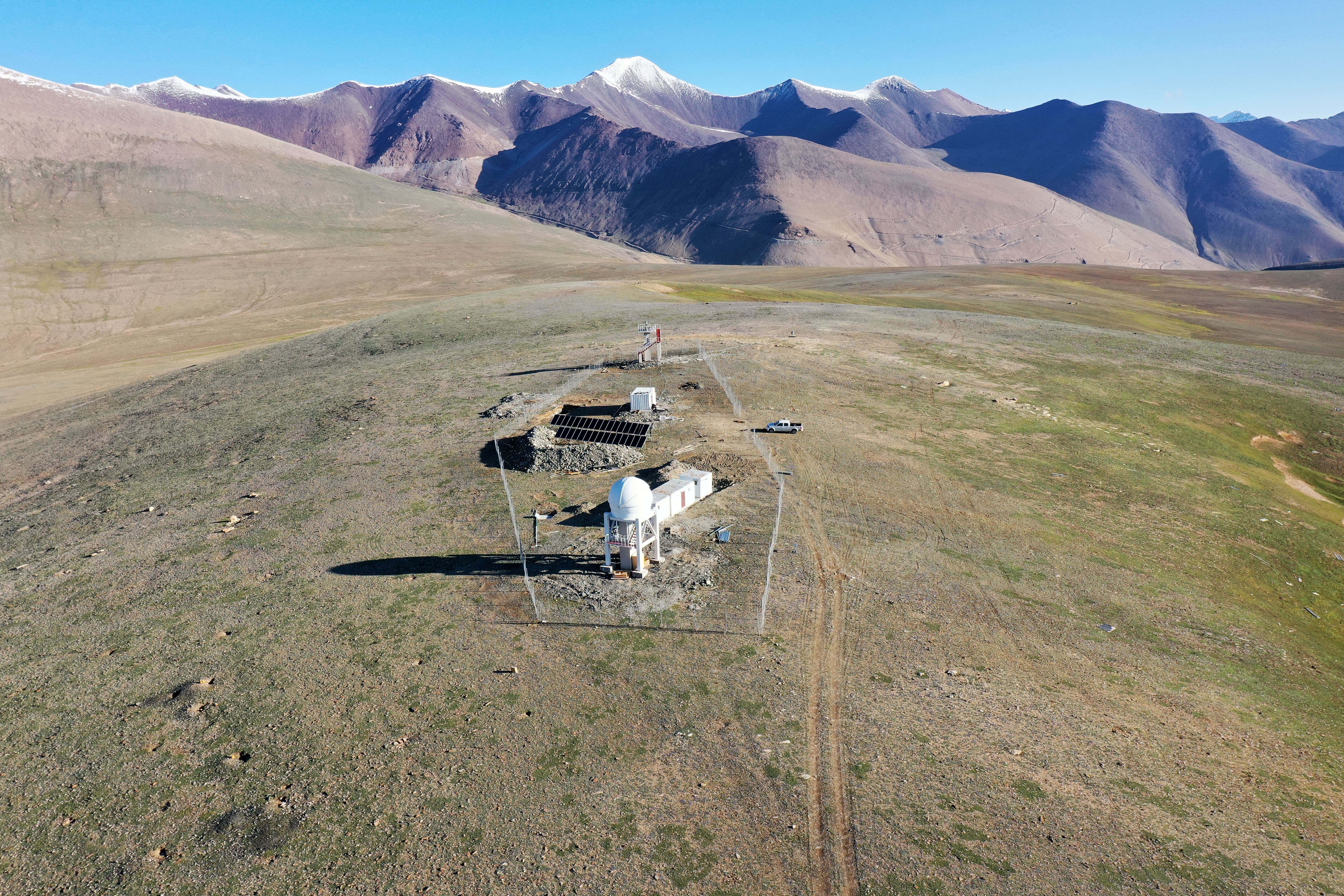}
   \caption{The newest general view of Muztagh-ata site.}
   \label{fig:gv}
\end{figure}

This article is the fifth of a series of papers reporting on the results of site-testing at Muztagh-ata site \citep{2020RAA....20...86X}. The geographical coordinates of this site are 38$^{\circ}$19'47"N, 74$^{\circ}$53'48"E, with an altitude of 4526 m above sea level, it is named for its proximity to the famous Muztagh-ata. The recent general view of the site can be seen in Figure~\ref{fig:gv}. In the last five-year on-site study of the characteristics of this site, we have shown the observing conditions monitoring results of Muztagh-ata site in previous work such as meteorological parameters, seeing, PWV et al \citep{2020RAA....20...87X,2022PASP..134a5006X}. This manuscript mainly presents the results of long-term monitoring of cloud amounts obtained from ground-based all-sky cameras from 2017 to 2021 at this site.

Cloud amount is the most straightforward parameter in estimating optical astronomical observatories because it is crucial for data quality and the observing time of the telescopes. Methods for measuring or analyzing cloud cover used in site-testing include satellite data analysis, telescope observation logs, and ground-based cloud detections as radio-meter \citep{2006JGRD..11111204L,2021JGRD..12634113B}, sky quality meter \citep{2020MNRAS.493.2463C,2022MNRAS.tmp.2789P} and all-sky cameras (ASC). Satellite data could support site selection in the initial stage but hardly provide a precise description of cloud amount for long-term measurements as the limitation of its resolution \citep{2012MNRAS.427.1903H,2015MNRAS.452.2185C,2020RAA....20...81C}. The observation logs of telescopes could provide very accurate statistical results of available observing time (hereafter AOT) for completed observatories \citep{2015PASP..127.1292Z,2016PASP..128j5003T,2020RAA....20..149X}. ASC can provide an accurate indication of statistical characteristics of cloud amount, so it is more widely used in initial stage site-testing for large optical telescope projects \citep{2006SPIE.6267E..2OW,2011RMxAC..41...36M} with its portability. 

In the site-testing campaign for China's future LOT (Large Optical Telescope) project, ASCs developed by \cite{Wang2020} have been used for cloud amount measurements at the three candidates and we have obtained some preliminary conclusion \citep{2020RAA....20...81C}. Manual checking is the main processing mode of all-sky images at our site until now. We adopt the method of cross-checking by multi persons to ensure the accuracy of the results. From 2017 January to 2021 November, we acquired $\sim$150,000 all-sky images at the Muztagh-ata site. Long-term statistical analysis of cloud cover is essential to assess the observing condition of our site. So we reviewed all the images to seek the time-varying characteristics of cloud cover in 2022. The structure of this paper is as follows: In Section 2, we give a brief introduction to the ASC and data acquisition. In Section 3, we provide the detailed statistical results of nighttime cloud amounts and the continuity of AOT in the last five years. Conclusions are given in Section 4.

\section{Cloud Amount Measurement}
\begin{figure}
   \centering
  \includegraphics[width=10cm, angle=0]{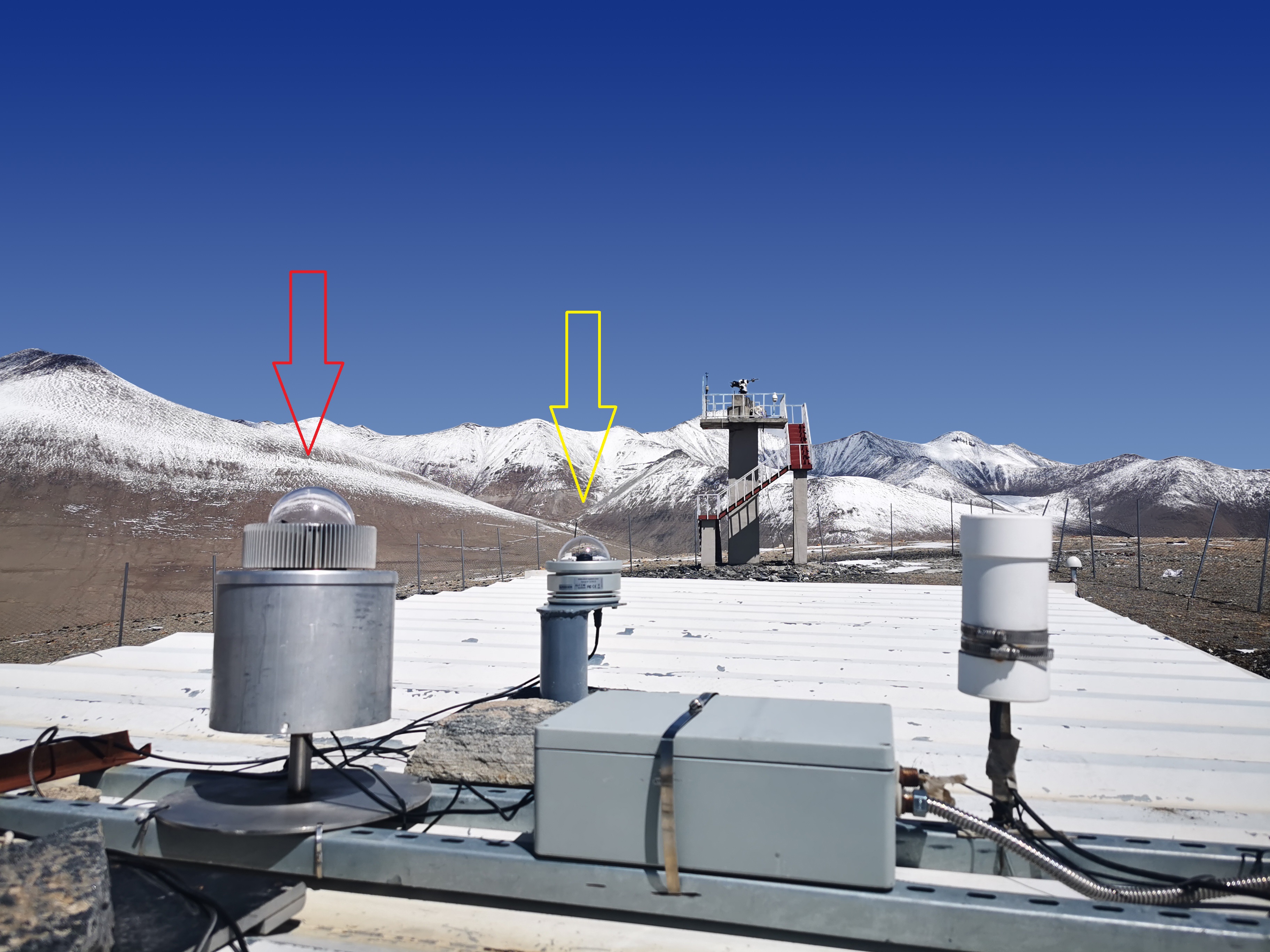}
   \caption{ASCs at Muztagh-ata site.}
   \label{fig:xj_1}
\end{figure}

The Muztagh-ata site is equipped with two ASCs as displayed in Figure~\ref{fig:xj_1}. However, the ASC marked by the yellow arrow is operational only since March 2022. This article thus mainly relies on data obtained with the other one marked by the red arrow. It is a fisheye lens consisting of a Canon Digital Single Lens Reflex Camera with 3456$\times$5184 pixels. The camera captures images at a 20-minute sampling interval during daytime while 5 minutes during nighttime, and with different exposure times depending on the sky background measured by sky brightness measurement (SBM). Here we define the time from the beginning of astronomical morning twilight to the end of astronomical evening twilight as daytime and the rest of this day as nighttime. 

The all-sky images have been classified into four categories following the Skidmore methodology \citep{2008SPIE.7012E..24S}: ``clear"(No cloud), ``outer"(No cloud within the inner circle, and cloud within the outer circle), ``inner"(No more than 50$\%$ cloud within both the inner and outer circle), ``covered"(Coverage of cloud within inner + outer circle over 50$\%$). Here, the sky area with zenith angles smaller than $44.7^\circ$ is defined as the inner circle, and the sky area with zenith angles between $44.7^\circ$ to $65^\circ$ is defined as the outer circle. An example of the four definitions of all-sky images can be seen in Figure 9 of \cite{2020RAA....20...82C}. 

\begin{figure*}
   \centering
  \includegraphics[width=14cm, angle=0]{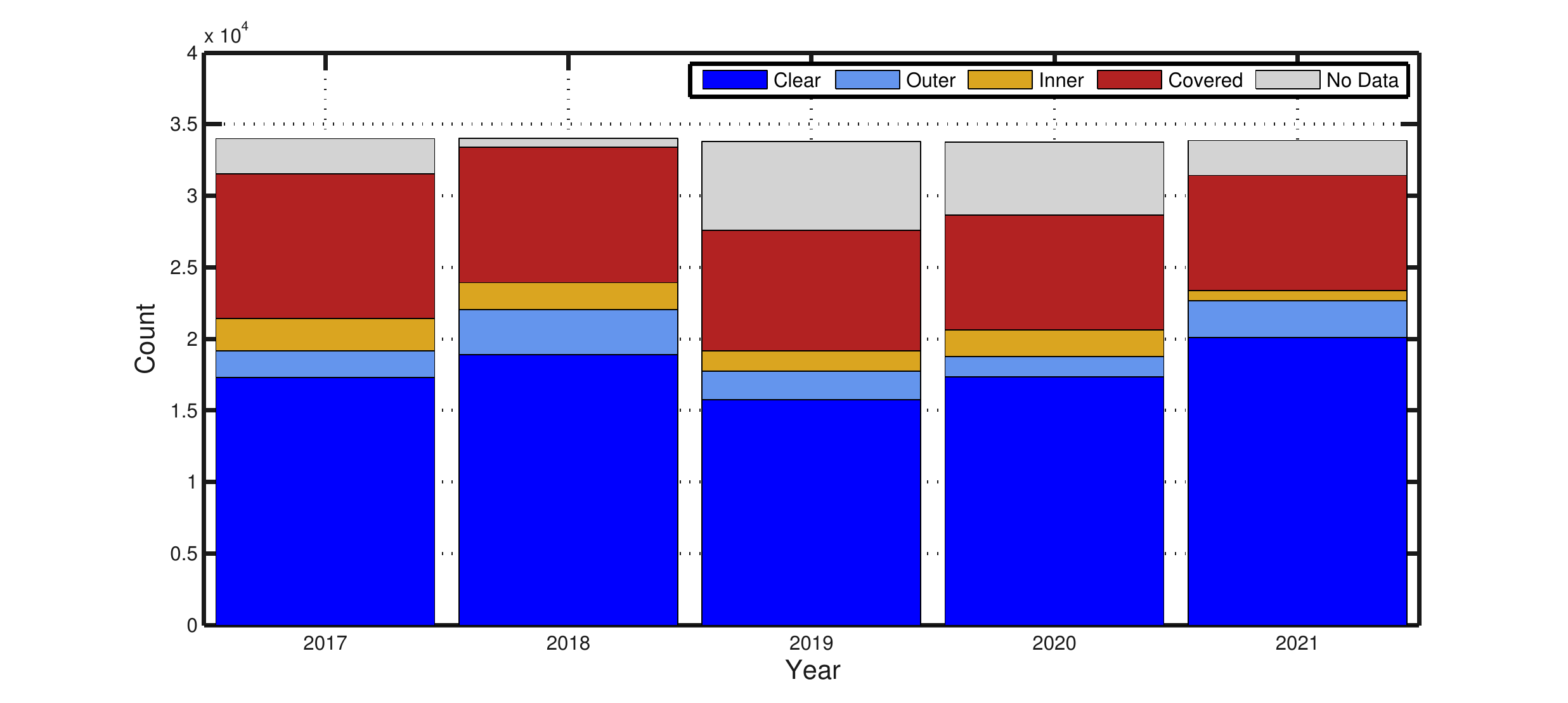}
   \caption{Total all-sky images of the last five years and the fraction of AOT.}
   \label{fig:xj_2}
\end{figure*}

\begin{figure*}
\centering
\includegraphics[width=14cm, height= 18cm,angle=0]{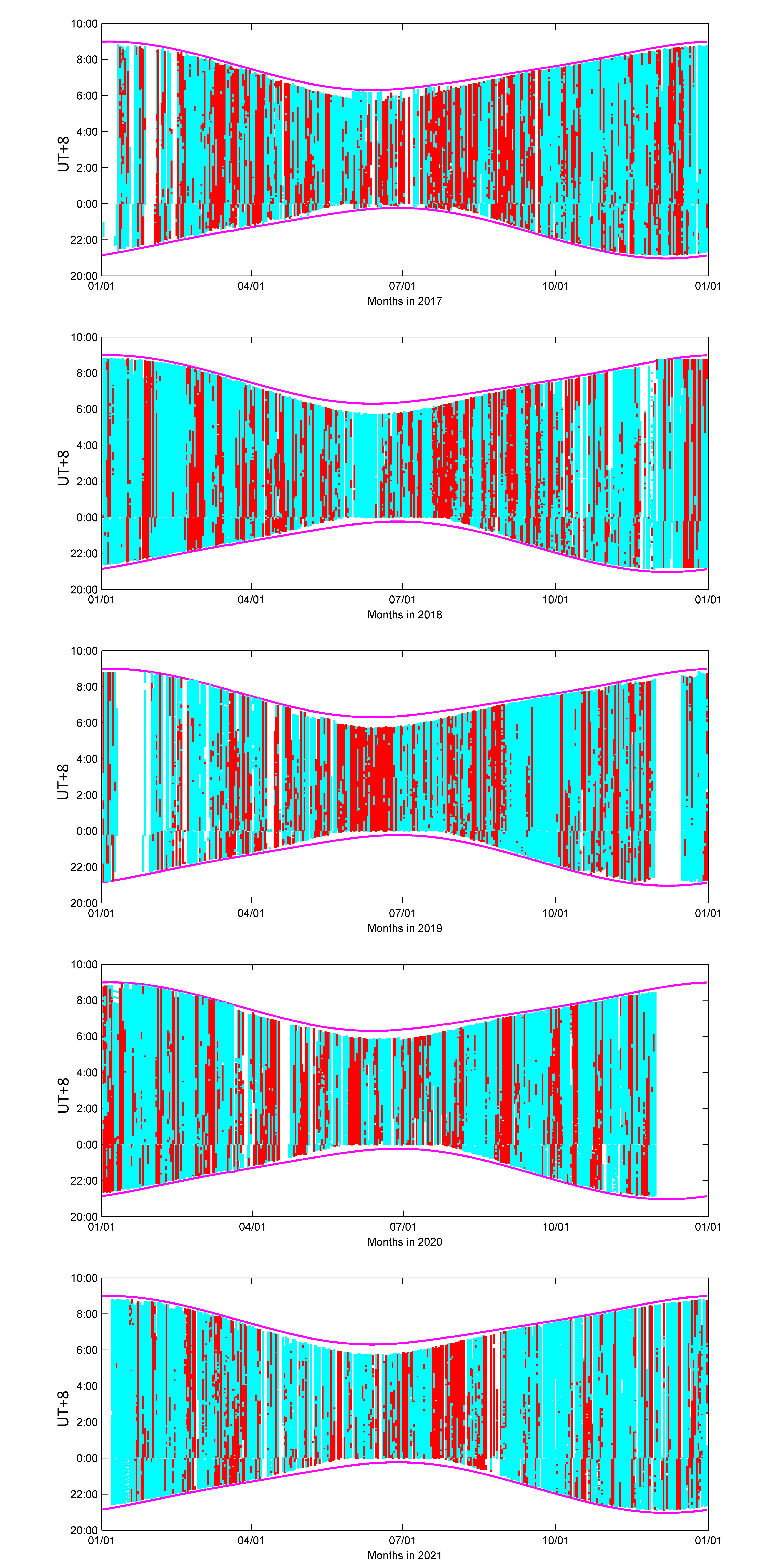}
\caption{Distribution of cloud cover from 2017 to 2021. The turquoise bars represent AOT and red bars represent UOT, and the pink curves represent twilights during the observation period.}
\label{fig:xj_3}
\end{figure*}

The ASC acquires $\sim$34,000 images during nighttime per year. We visually inspected every image and defined its classification according to the distribution of clouds and stars. In Figure~\ref{fig:xj_2} we present the number of images obtained each year from 2017 to 2021, blue for ``clear", light blue for ``outer", orange for ``inner", and red for ``covered". The gray parts in the bars represent the missing data. The distributions of night cloud cover at the Muztagh-ata site in the last five years are plotted in Figure~\ref{fig:xj_3}, the turquoise points in the plots represent the cloud amount as ``clear" or ``outer" (we define these two kinds of images as AOT), and red for cloud amount as ``inner" or ``covered" (we define these two kinds of images as unavailable observing time, hereafter UOT). Data acquisition for the five years is listed in the last column of Table~\ref{tab:xj_1}. In 2019, the data acquisition was only 81.1\%. There were two gaps both lasting about one month in that year: a power failure in January, and the two cameras were sent back to Urumqi for repair in November. Besides, At the beginning of 2017 and the end of 2020, there were long data interruptions due to unstable power.

\section{Cloud Amount Statistics }
Table~\ref{tab:xj_1} shows the yearly statistics of the classification of nighttime all-sky images at the Muztagh-ata site for the last five years. The percentage of AOT of the whole nighttime data was 65\%. There is a difference in yearly nighttime cloud amounts. The lowest fraction of AOT during nighttime is 61\% in 2017, and the highest is 72\% in 2021. It should be noted that the data acquisition was lower than 85\% in 2019 and 2020, the actual fraction of AOT time may be slightly different.

\begin{table}[htb]
    \centering
      \caption{Statistics of the classification of nighttime all-sky images and data acquisition.}
\label{tab:xj_1} 
\begin{tabular}{cccccc}
\hline
Year                 & Clear & Outer & Inner & Covered  & Data Acquisition \\
\hline
2017                 & 55\%    &  6\%    & 7\%    & 32\%  &   92.7\%\\
2018                 & 56\%    &  10\%   & 5\%    & 29\%  &    98.2\%\\
2019                 & 57\%    &  7\%    & 6\%    & 30\%  &   81.1\%\\
2020                 & 60\%    &  5\%    & 6\%    & 29\%  &   84.3\%\\
2021                 & 64\%    &  8\%    & 2\%    & 26\%  &   92.4\%\\
\hline
Total                & 58\%    &  7\%    & 6\%    & 29\%   & 89.7\% \\
\hline
\end{tabular}
\end{table}

\begin{figure}
   \centering
  \includegraphics[width=16cm, angle=0]{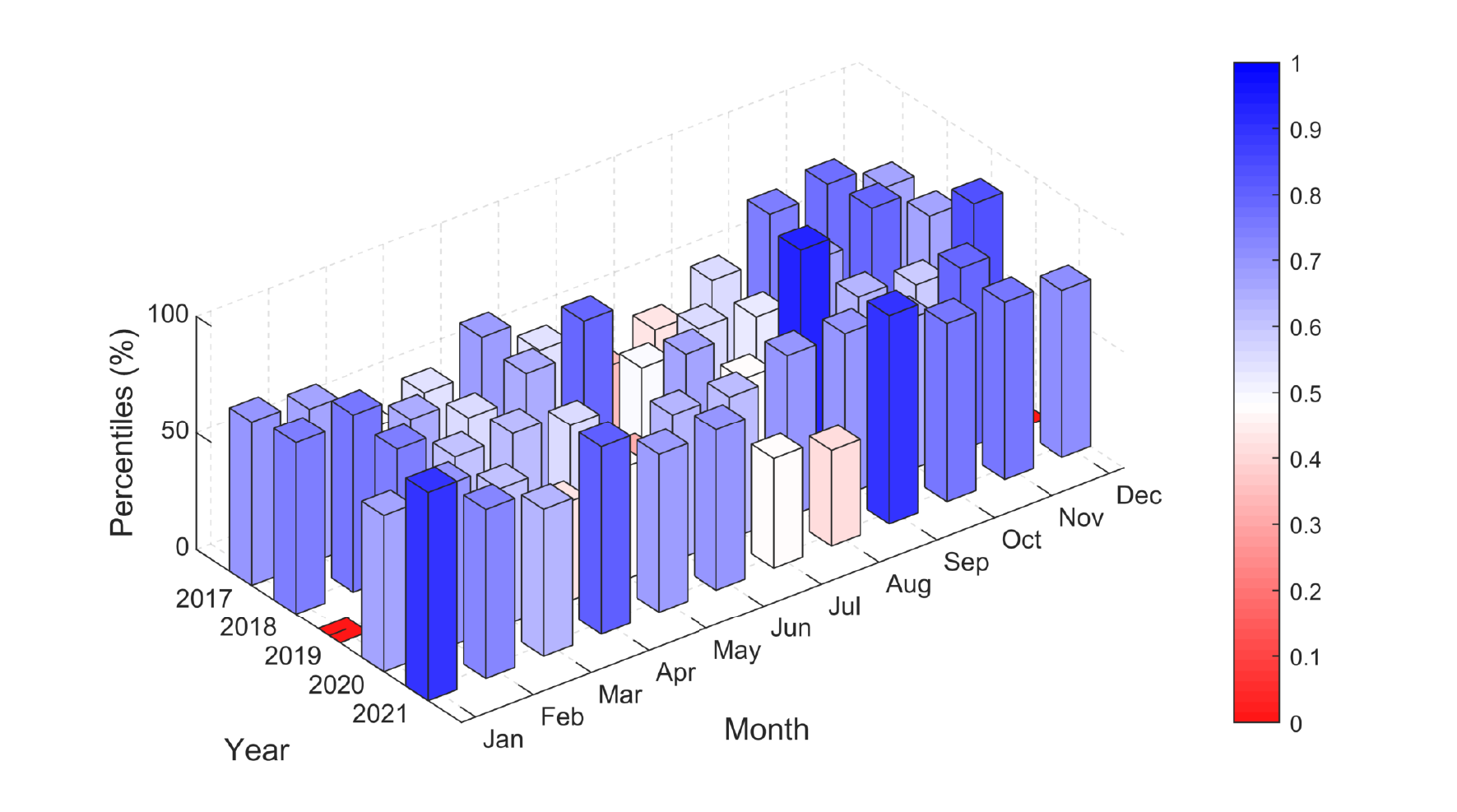}
   \caption{The percentiles of AOT within times when data was acquired during the nights of each month from 2017 to 2021.}
   \label{fig:xj_4}
\end{figure}

Figure~\ref{fig:xj_4} depicts monthly statistics of the percentiles of AOT within times when data was acquired during the nights for the five years. The nighttime cloud amount in the period from April to August is higher than that in the remaining months. There was one month in the warm seasons of all the years with an AOT fraction as low as 40\%. The month with the most cloud cover was June of 2019, the proportion of AOT in this month was 32$\%$. The month with the least cloud cover was September of the same year, and the proportion of AOT was 93\%.

Previous works have explored different standards to recognize observing nights. \cite{2016PASP..128j5003T} calculated the observing nights of the Delingha station as the sum of clear nights and cloudy nights. \cite{2016PASP..128j5003T} adopted the definition of the clear night used by \cite{1973PASP...85..255M} and the cloudy night used by \cite{1992RMxAA..24..179T}. A clear night means the night is clear for at least 6 hours, and a cloudy night means a night with a sky coverage of less than 65\%. In this paper, we consider an observing night as a night with more than 50\% AOT and a clear night as a night with more than 75\% of the AOT. 

\begin{figure}
   \centering
  \includegraphics[width=14cm, angle=0]{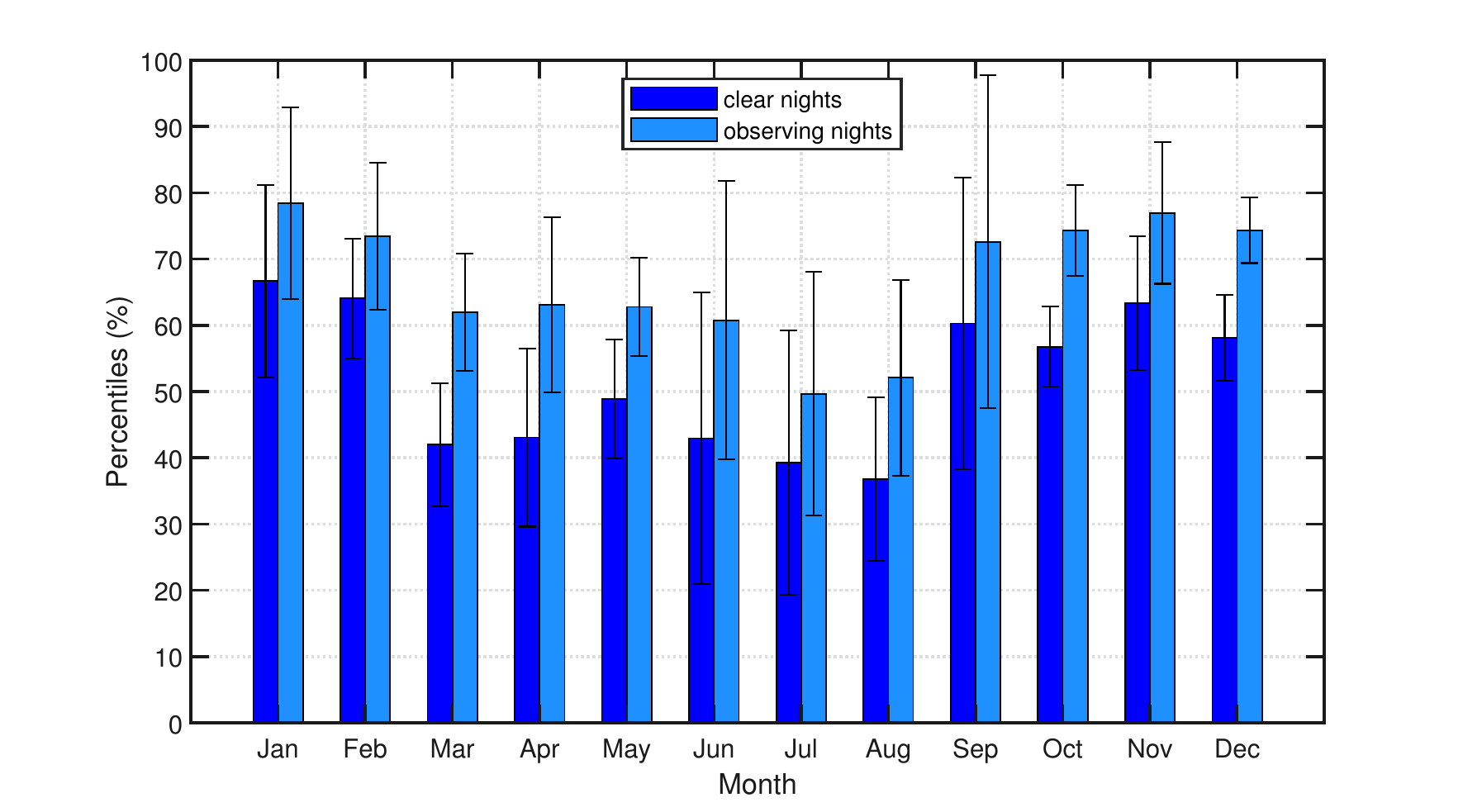}
   \caption{Monthly mean percentiles of the number of observing and clear nights within times when data was acquired from 2017 to 2021. Navy and light blue bars represent the monthly mean proportion of clear and observing nights respectively.}
   \label{fig:xj_5}
\end{figure}

Figure~\ref{fig:xj_5} shows the monthly percentiles of the number of observable nights and clear nights within times when data was acquired obtained by the acquired all-sky images. July had the least observing nights, and the proportion was 50\%. In the period from September to February, more than 70\% of nights were observing nights. Monthly statistics of observing nights and clear nights during the five years are expressed in Table~\ref{tab:xj_2}. There were long-time data gaps that occurred in January 2017, January and December 2019, and December 2020. It will affect the accuracy of estimating the yearly mean observing and clear nights. We used the mean value of the same period data of other years to replace the missing data during the calculation process and presented them with $\ast$ in Table~\ref{tab:xj_2}. In 2021, there were in total of 248 observing nights, and among them 192 clear nights. The mean values of observing and clear nights were 227 and 175 in the five years.

\begin{table*}[htb]
    \centering
      \caption{Amount of observing and clear nights from 2017 to 2021.}
\label{tab:xj_2} 
\begin{threeparttable}
\begin{tabular}{ccccccccccccccc}
\hline\hline
\multicolumn{1}{c}{Year} &  \multicolumn{2}{c}{2017(days)} & & \multicolumn{2}{c}{2018(days)} &&\multicolumn{2}{c}{2019(days)} &&\multicolumn{2}{c}{2020(days)}& &\multicolumn{2}{c}{2021(days)} \\
\cline{2-3}\cline{5-6}\cline{8-9}\cline{11-12}\cline{14-15}
Month            &$\ge$75\% &$\ge$50\%&  &$\ge$75\% &$\ge$50\%&  &$\ge$75\% &$\ge$50\%&  &$\ge$75\% &$\ge$50\%&  &$\ge$75\% &$\ge$50\%   \\
\hline
1	&	13$\ast$	&	25$\ast$	&	&	21	&	25	&	&	19$\ast$	&	24$\ast$	&	&	18	&	21	&	&	22	&	24	\\
2	&	11	&	14	&	&	18	&	20	&	&	16	&	18	&	&	19	&	22	&	&	18	&	20	\\
3	&	9	&	15	&	&	15	&	21	&	&	15	&	20	&	&	12	&	17	&	&	12	&	20	\\
4	&	11	&	18	&	&	10	&	17	&	&	12	&	16	&	&	6	&	10	&	&	17	&	21	\\
5	&	17	&	22	&	&	16	&	19	&	&	12	&	14	&	&	7	&	11	&	&	15	&	20	\\
6	&	8	&	15	&	&	19	&	22	&	&	3	&	6	&	&	14	&	18	&	&	14	&	21	\\
7	&	4	&	6	&	&	10	&	14	&	&	18	&	21	&	&	14	&	18	&	&	11	&	13	\\
8	&	8	&	13	&	&	10	&	14	&	&	10	&	14	&	&	19	&	22	&	&	6	&	12	\\
9	&	10	&	13	&	&	11	&	16	&	&	26	&	29	&	&	17	&	22	&	&	24	&	26	\\
10	&	18	&	24	&	&	14	&	20	&	&	15	&	22	&	&	17	&	19	&	&	20	&	25	\\
11	&	19	&	26	&	&	16	&	17	&	&	15	&	18	&	&	19	&	23	&	&	19	&	23	\\
12	&	16	&	19	&	&	18	&	21	&	&	21$\ast$	&	23$\ast$	&	&	16$\ast$	&	21$\ast$	&	&	14	&	23	\\
\hline
Total & 	144	&	210	&	&	178	&	226	&	&	182	&	225	&	&	178	&	224	&	&	192	&	248	\\
\hline
\end{tabular}

\footnotesize{The data with $\ast$ means it was estimated by the same period data of other years.}
\end{threeparttable}
\end{table*}

Limited to the monitoring method, all-sky images used to measure cloud amounts are not sensitive to the high cloud in the visible band. Satellite observations are useful validation and supplements for in situ measurement. We used CLARA (CM SAF cLoud, Albedo and surface RAdiation) dataset \citep{2017TCDR,2021ICDR,2021EGUGA..23.8167T} provided by the EUMETSAT Satellite Application Facility on Climate Monitoring\footnote{https://cds.climate.copernicus.eu/cdsapp$\#$!/home}, to compare with our in situ results. This remote sensing dataset includes the daily and monthly means of cloud fractional cover, cloud top level, and other cloud physical properties with a horizontal resolution of 0.25$^\circ$$\times$ 0.25$^\circ$ from 1982 to now. 

\begin{figure}
   \centering
  \includegraphics[width=14cm, angle=0]{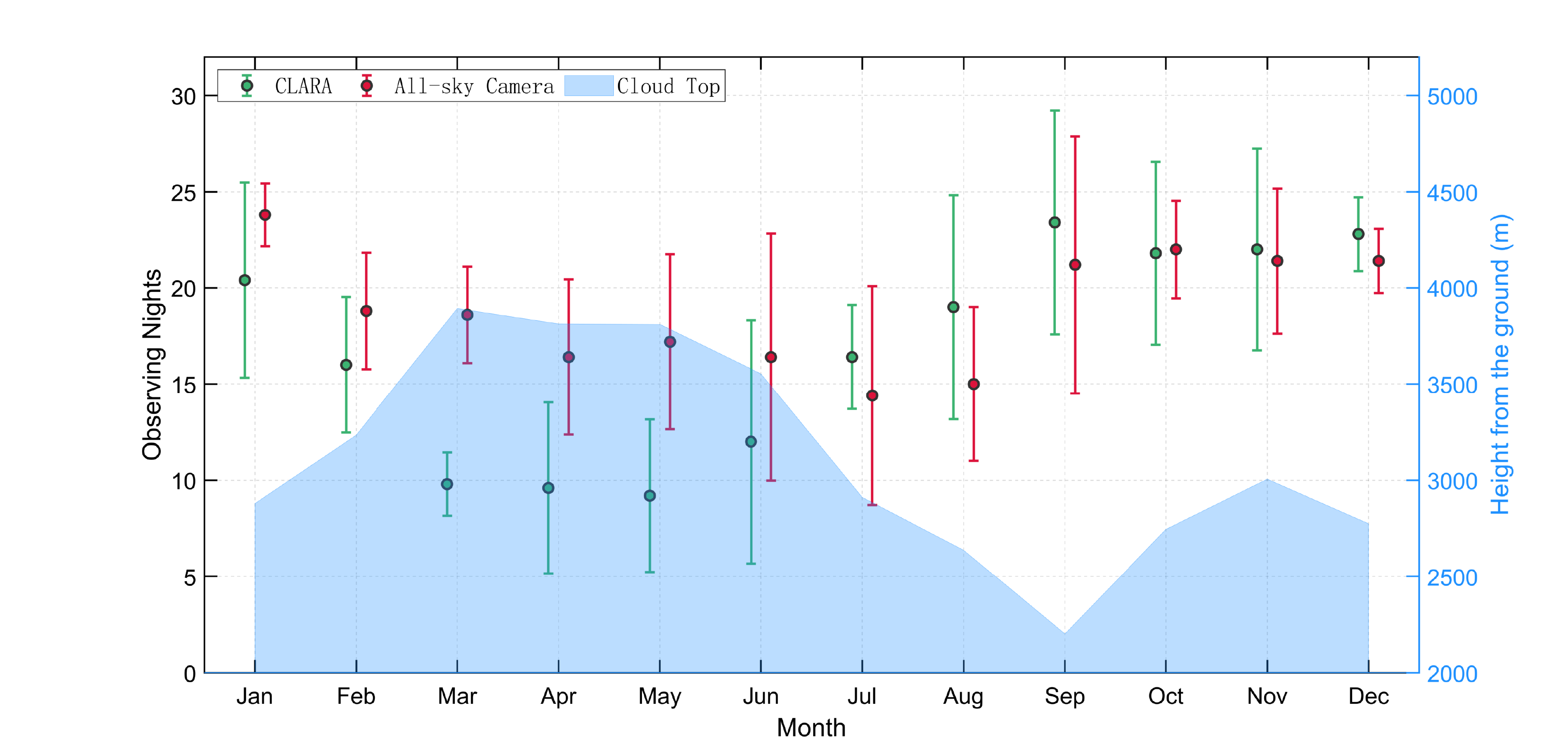}
   \caption{Comparison between our all-sky camera and CLARA product. Red and green circles represent the monthly number of observing night results from the all-sky cameras and CLARA respectively, and the blue area represents the cloud top heights, which are measured from the ground topography in meters.}
   \label{fig:compare}
\end{figure}

We calculated the monthly mean observing nights for our site using the CLARA dataset and ground-based measurements from 2017 to 2021, and the results are depicted in Figure~\ref{fig:compare}. The blue shaded area in Figure~\ref{fig:compare} represents the monthly mean value of the cloud top height (measured from the ground topography in meters). After some trials, we defined one night with a cloud fraction less than 60\% as an observable night when using ICDR. The two data results are consistent well in most months. But in the period from March to June, there are obviously differences. Obtained from CLARA, the yearly mean amount of observing nights was 202. 

The consistency of the cloud measurement between the satellite and ASC data is affected by many factors\citep{2020RAA....20...82C}, such as the observation band, the field of view, the processing algorithm\citep{2022MNRAS.511.5363W}, and the different detection capability for clouds with different heights. In the months from March to July, the heights of cloud tops were obviously higher than in the other months, and the discrepancy in observing nights between satellite data and our all-sky images was larger than in other months. It may be caused by the insensitivity to the higher cloud for our visual band ground-based all-sky camera and one of the reasons for the higher statistic of cloud fraction, correspondingly lower observing nights, by remote sensing dataset during these months.

\begin{figure*}
   \centering
  \includegraphics[width=10cm, angle=0]{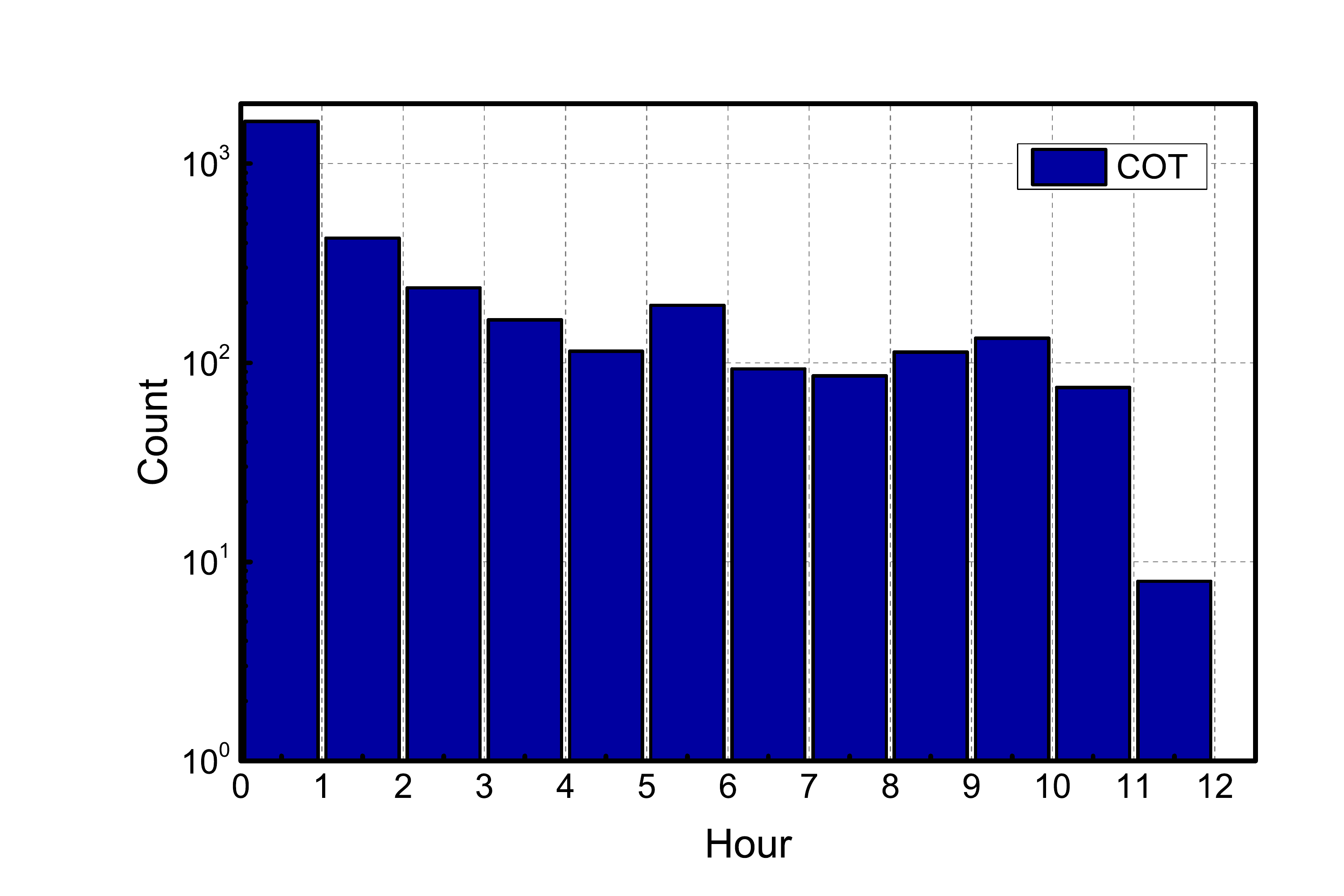}
   \caption{Statistic of COT for different lengths of all data.}
   \label{fig:xj_6}
\end{figure*}

\begin{figure}
   \centering
  \includegraphics[height=8cm, angle=0]{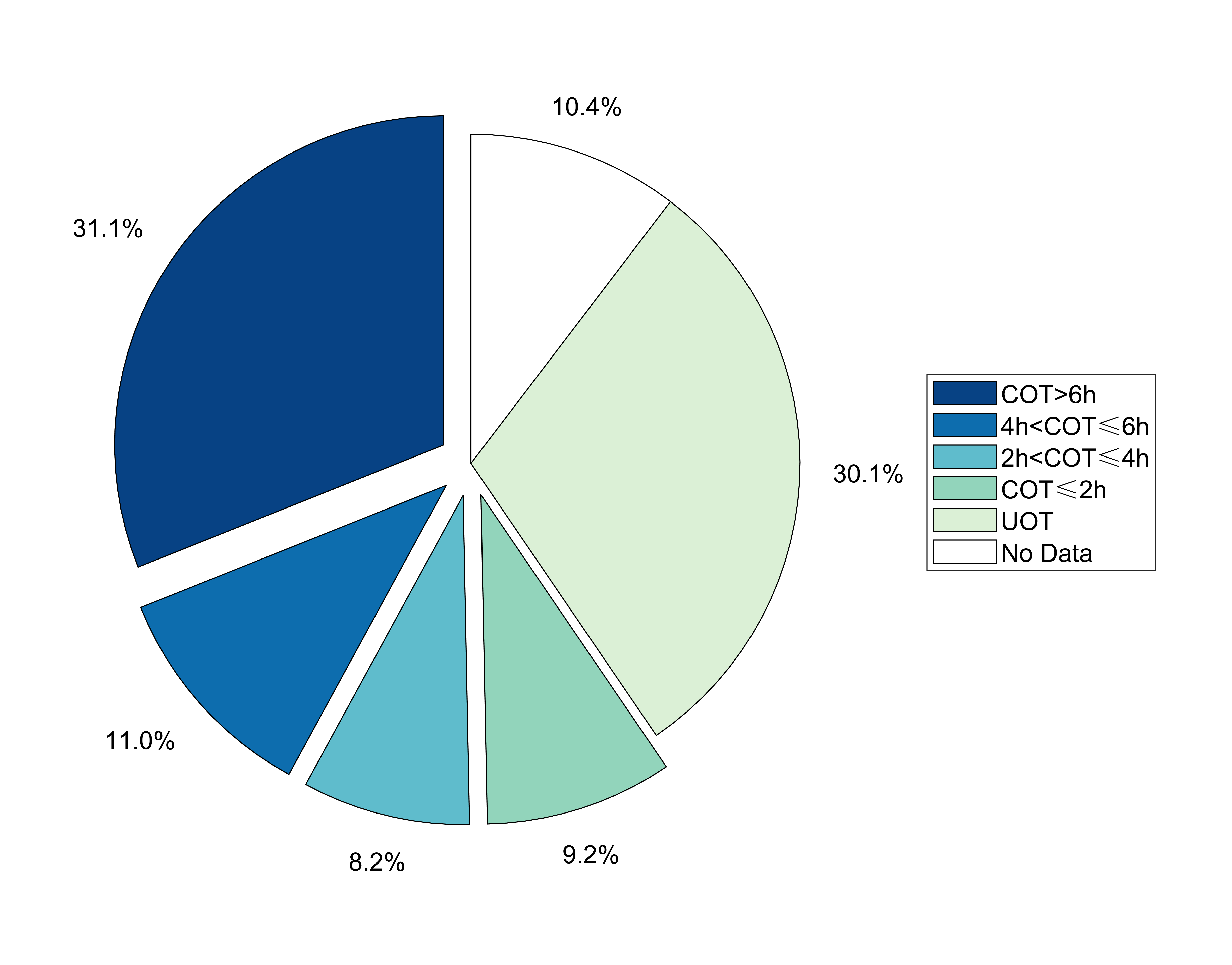}
   \caption{Percentiles of the four ranges of COT in the nighttime.}
   \label{fig:xj_7}
\end{figure}

In many observations, time-domain astronomy for example, continuous observing time (hereafter COT) is very important. Interruptions due to passing clouds will cause loss of useful information. The statistics of the COT of an astronomical observatory are crucial for evaluating these observations whether can be conducted efficiently. We studied the distribution of COT at the Muztagh-ata site, histogram plotted in Figure~\ref{fig:xj_6} presents the amount of COT for different time lengths of the whole dataset. The count of COT longer than 6 hours is 508 for all image data and on average more than 100 for each year. We divided all the COT into four ranges: longer than 6 hours, between 4 and 6 hours, between 2 and 4 hours, and less than 2 hours. The percentiles of the four COT ranges in all nighttime are shown in Figure~\ref{fig:xj_7}. Except for 10.4\% data missing and 30.1\% unobserving, the observing time with COT of more than 6 hours accounts for 31.1\% of the total nighttime, 11.0\% for the range of 4 to 6 hours, 8.2\% for 2 to 4 hours, and 9.2\% for less than 2 hours. The mean numbers of nights with COT over 4 hours for every month from 2017 to 2021 are listed in Table~\ref{tab:xj_3}. There are in total of 169 nights that observation can be conducted continuously for at least four hours every year.

\begin{table*}
    \centering
      \caption{Monthly statistics of nights with COT longer than 4 hours.}
\label{tab:xj_3} 
\begin{tabular}{ccccccc}
\hline
month & 2017(days) & 2018(days)  & 2019(days)  & 2020(days)  & 2021(days)  & mean(days)\\
\hline
1	&	11	&	23	&	-	&	19	&	24	& 19\\
2	&	12	&	20	&	13	&	18	&	19	&16\\
3	&	9	&	17	&	10	&	11	&	12	&12\\
4	&	13	&	12	&	8	&	8	&	14	&11\\
5	&	18	&	14	&	10	&	9	&	12	&13\\
6	&	6	&	18	&	3	&	11	&	8	&9\\
7	&	4	&	7	&	13	&	13	&	9	&9\\
8	&	8	&	9	&	7	&	17	&	4	&9\\
9	&	12	&	10	&	26	&	15	&	25	&17\\
10	&	18	&	12	&	18	&	18	&	20	&17\\
11	&	21	&	16	&	16	&	22	&	19	&19\\
12	&	21	&	22	&	13	&	-	&	15	&18\\
\hline
Total    &	153	&	180	&	137	&	161	&	181	&169\\            
\hline
\end{tabular}
\end{table*}

\section{Conclusion}
In the last five years, we have used all-sky cameras for cloud amount measurement at the Muztagh-ata site and have processed all the ASC image data to obtain the cloud cover distribution results for the five years. The main results at our site are as follows:

1. Statistics results from ground-based monitor show that the average proportion of AOT during nighttime was 65\% during the last five years. The yearly proportion of observing and clear nights was 66\% and 51\%, respectively. We can estimate that the Muztagh-ata site can provide over 227 observing nights, 175 clear nights, and 169 nights with COT over 4 hours every year.

2. From 2017 to 2021, there is some difference in the yearly mean cloud coverage at the Muztagh-ata site. The proportion of AOT during nighttime was 61\% in 2017 while 72\% in 2021. The discrepancy in the amount of observing nights between 2019 and 2021 was 30, which get from the CLARA ICDR dataset. More extended measurements for cloud cover are needed to confirm whether the yearly difference in cloud coverage is an inherent cyclical characteristic of the local climate. Monthly cloud cover distribution shows strong seasonal dependence. The best months in clarity conditions are September to February of the following year.

3. Monthly statistics of observing nights from our in situ measurement and satellite observation are consistent well in the months with average cloud top heights lower than 3500m from the ground. We believe that the middle or high cloud is the main factor affecting the results. It needs more detailed research to improve the ability to recognize high clouds for our in situ measurements. An infrared cloud camera is about to be equipped at our site.

\begin{acknowledgements}
This work was supported by the Chinese Academy of Science (CAS) "Light of West China" Program (No.2022$\_$XBQNXZ$\_$014), the Xinjiang Natural Science Foundation (Grant No. 2022D01A357), the Joint Research Fund in Astronomy under a cooperative agreement between the National Natural Science Foundation of China (NSFC) and the Chinese Academy of Sciences (CAS) (grant No.U2031209), the National Natural Science Foundation of China (Grant No.11873081, 11603065, and 12073047), resource sharing platform construction project of Xinjiang Uygur Autonomous Region (No.PT2306).

\end{acknowledgements}


\end{document}